\documentclass[letterpaper, 9 pt, conference]{ieeeconf}
\IEEEoverridecommandlockouts 
\IEEEoverridecommandlockouts 
\usepackage{amsmath,amssymb,latexsym}
\usepackage[dvips]{graphicx}
\usepackage{graphicx}
\usepackage{psfrag}
\usepackage{epsfig}
\usepackage{float}
\usepackage{array}
\usepackage{mathrsfs}
\usepackage[geometry]{ifsym}
\usepackage{color}
\usepackage{fancyhdr}
\usepackage{subfigure}
\newtheorem{thm}{Theorem}
\newtheorem{assumption}{Assumption}
\newtheorem{remark}{Remark}
\title{\bf Multi-Parametric Extremum Seeking-based Auto-Tuning for Robust Input-Output Linearization Control}

\author{ Mouhacine Benosman
\thanks{Mouhacine Benosman (m{\_}benosman@ieee.org) is with Mitsubishi Electric Research Laboratories, 201 Broadway Street, Cambridge, MA 02139, USA. Accepted at the IEEE CDC 2015}
 }

\date{}

\begin{document}

\maketitle

\begin{abstract}
We study in this paper the problem of iterative feedback gains
tuning for a class of nonlinear systems. We consider Input-Output
linearizable nonlinear systems with additive uncertainties. We
first design a nominal Input-Output linearization-based controller
that ensures global uniform boundedness of  the output tracking
error dynamics. Then, we complement the robust controller with a
model-free {\it multi-parametric extremum seeking} (MES) control
to iteratively auto-tune the feedback gains. We analyze the
stability of the whole controller, i.e. robust nonlinear
controller plus model-free learning algorithm. We use numerical
tests to demonstrate the performance of this method on a
mechatronics example.
\end{abstract}
\section{Introduction}
Input-Output feedback linearization with static state feedback is
a very well known nonlinear control approach, which has been
extensively used to solve trajectory tracking for  nonlinear
systems \cite{I89}. Its robust version has also been extensively
studied, e.g. \cite{FK08,CW95,PD11,FBPG06}. The main approaches
proposed, either combine a linear robust controller with the
linearization controller to achieve some robustness w.r.t. to
structural model uncertainties and measurable disturbances, e.g.
\cite{PD11} and references therein, or use high gains observers to
estimate the input disturbance and use the estimation to
compensate for the disturbance and recover some performance of the
feedback linearization controller, e.g.\cite{FK08}. In this work
we focus on specific problem for Input-Output feedback
linearization control, namely, iterative
feedback gains tuning. \\
Indeed, the use of learning algorithm to tune feedback gains of
nominal linear controllers to achieve some desired performances
has been studied in several papers, e.g.
\cite{LGMBT03,H02,KK06,KRK06}. In this work, we try to extend
these approaches to a more general setting of uncertain nonlinear
systems (refer to \cite{benosmanatincacc13} for preliminary
results). We consider here a particular class of nonlinear
systems, namely, nonlinear models affine in the control input,
which are linearizable via static state feedback. We consider
bounded additive model uncertainties with known upper bound
function. We propose a simple modular iterative gains tuning
controller, in the sense that we first design a passive robust
controller, based on the classical Input-Output linearization
method merged with a Lyapunov reconstruction-based control, e.g.
\cite{K96,BL09-4}. This passive robust controller ensures uniform
boundedness of the tracking errors and their convergence to a
given invariant set. Next, in a second phase we add a
multi-variable extremum seeking algorithm to iteratively auto-tune
the feedback gains of the passive robust controller to optimize a
desired system performance, which is formulated in terms of a
desired cost function minimization.\\
This paper is organized as follows: First, some notations and
definitions are recalled in Section \ref{prem}. Next, we present
the class of systems studied here and formulate the control
problem in Section \ref{problem_formulation}. The proposed control
approach together with the closed-loop dynamic solutions
boundedness are presented in Section \ref{control_design}. Section
\ref{example} is dedicated to the application of the controller to
a mechatronics example. Finally the paper ends with a summarizing
conclusion in Section \ref{conclusion}.

\section{Notations and definitions}\label{prem}

Throughout the paper we will use $|.|$ to denote the Euclidean
norm; i.e., for $x\in \mathbb{R}^n$ we have $|x|=\sqrt{x^T x}$. We
will use the notations $diag\{m_{1},...,m_{n}\}$ for $n\times n$
diagonal matrix, $z(i)$ denotes the $i$th element of the vector
$z$. We use $\dot{ (.)}$ for the short notation of time derivative
and $f^{(r)}(t)$ for $\frac{d^{r}f(t)}{dt^{r}}$. $Max(V)$ denotes
the maximum element of a vector $V$, and $sgn(.)$ denotes for the
sign function. We denote by $\mathbb{C}^k$ functions that are $k$
times differentiable, and by $\mathbb{C}^{\infty}$ a smooth
function. A function is said analytic in a given set, if it admits
a convergent Taylor series approximation in some neighborhood of
every point of the set. An impulsive dynamical system is said to
be well-posed if it has well defined distinct resetting times,
admits a unique solution over a finite forward time interval and
does not exhibits any Zeno solutions, i.e. an infinitely many
resetting of the system in finite time interval \cite{HCN06}.
Finally, in the sequel when we talk about error trajectories
boundedness, we mean uniform boundedness as defined in \cite{K96}
(p.167, Definition 4.6 ) for nonlinear continuous systems, and in
\cite{HCN06} (p. 67, Definition 2.12) for time-dependent impulsive
dynamical systems.
\section{Problem formulation}\label{problem_formulation}
\subsection{Class of systems}
We consider here affine uncertain nonlinear systems of the form:
\begin{equation}\label{systemeaffine1}
\begin{array}{c}
\dot x=f(x)+\Delta f(x)+g(x)u,\;x(0)=x_{0}\\
y=h(x),
\end{array}
\end{equation}
where
$x\in\mathbb{R}^{n},u\in\mathbb{R}^{n_{a}},y\in\mathbb{R}^{m}\;(n_{a}\geq
m)$, represent respectively the state, the input and the
controlled output vectors, $x_{0}$ is a known initial condition,
$\Delta f(x)$ is a vector field representing additive model
uncertainties. The vector fields $f$, $\Delta f$, columns of $g$
and function $h$ satisfy the following assumptions.
\begin{assumption} $f: \mathbb{R}^{n} \rightarrow
\mathbb{R}^{n}$ and the columns of $g: \mathbb{R}^{n}\rightarrow
\mathbb{R}^{n\times n_{a}}$ are $\mathbb{C}^{\infty}$ vector
fields on a bounded set $X$ of $\mathbb{R}^{n}$ and $h(x)$ is a
$\mathbb{C}^{\infty}$ function on $X$. The vector field $\Delta
f(x)$ is $\mathbb{C}^{1}$ on $X$.
\end{assumption}
\begin{assumption} System (\ref{systemeaffine1}) has a
well-defined (vector) relative degree $\{r1,\dots,rm\}$ at each
point $x^0\in X$, and the system is linearizable, i.e.
$\sum_{i=1}^{i=m}r{i}=n$ (see e.g. \cite{I89}).
\end{assumption}
\begin{assumption} The uncertainty vector $\Delta f$ is
s.t. $|\Delta f(x)|\leq d(x)\;\forall x\in X$, where
$d\;:\;X\rightarrow  \mathbb{R}$ is a smooth nonnegative function.
\end{assumption}
\begin{assumption}\label{traj_assumption}
The desired output trajectories $y_{id}$ are smooth functions of
time, relating desired initial points $y_{i0}$ at $t=0$ to desired
final points $y_{if}$ at $t=t_{f}$, and s.t.
$y_{id}(t)=y_{if},\;\forall t\geq t_{f},\;t_{f}>0$,
$i\in\{1,...,m\}$.
\end{assumption}
\subsection{Control objectives}
Our objective is to design a feedback controller $u(x,K)$, which
ensures for the uncertain model (\ref{systemeaffine1}) uniform
boundedness of a tracking error, and for which the stabilizing
feedback gains vector $K$ is iteratively auto-tuned, to optimize a
desired performance cost function.
\\ We stress here
that the goal of the gain auto-tuning is not stabilization but
rather performance optimization. To achieve this control
objective, we proceed as follows: We design a `passive' robust
controller which ensures boundedness of the tracking error
dynamics, and we combine it with a model-free learning algorithm
to iteratively (resting from the same initial condition at each
iteration) auto-tune the feedback gains of the controller, and
optimize online a desired performance cost function.
\section{Controller design}\label{control_design}
\subsection{Step one: Passive robust control design}
Under Assumption 2 and nominal conditions, i.e. $\Delta f=0$,
system (\ref{systemeaffine1}) can be written as \cite{I89}:
\begin{equation}\label{sys-lin}
\begin{array}{l}
y^{(r)}(t)=b(\xi(t))+A(\xi(t))u(t),\\
\end{array}
\end{equation}
where
\begin{equation}\label{diffeo}
\begin{array}{l}
y^{(r)}(t)\triangleq (y_1^{(r_1)}(t),\dots,y_m^{(r_m)}(t))^T,\\
\xi(t)=(\xi^1(t),\dots,\xi^m(t))^T,\\
\xi^i(t)=(y_{i}(t),\dots,y_{i}^{(ri-1)}(t)),\;\;1\leq i\leq m,
\end{array}
\end{equation}
and $b,A$ write as functions of $f,g,h$, and $A$ is non-singular
in $X$ (\cite{I89}, pp. 234-288).\\At this point we introduce one
more assumption on the system.
\begin{assumption}\label{additive_uncer} We assume that the additive uncertainties
$\Delta f$ in (\ref{systemeaffine1}) appear as additive
uncertainties in the linearized model (\ref{sys-lin}),
(\ref{diffeo}), as follows
\begin{equation}\label{perturbed-model}
\begin{array}{l}
y^{(r)}=b(\xi)+\Delta b(\xi)+A(\xi)u,\\
\end{array}
\end{equation}
where $\Delta b$ is $\mathbb{C}^{1}$ on $\tilde{X}$, and s.t.
$|\Delta b(\xi)|\leq d_{2}(\xi)\;\forall \xi\in \tilde{X}$, where
$d_{2}\;:\;\tilde{X}\rightarrow  \mathbb{R}$ is a smooth
nonnegative function, and $\tilde{X}$ is the image of the set $X$
by the diffeomorphism $x\rightarrow \xi$ between the states of
(\ref{systemeaffine1}) and (\ref{sys-lin}).
\end{assumption}
\begin{remark}
Assumption \ref{additive_uncer}, can be ensured under the
so-called "matching conditions" (\cite{EO92}, p. 146).
\end{remark}
If we consider the nominal model (\ref{sys-lin}) first, we can
define a virtual input vector $v$ as
\begin{equation}\label{virtual}
b(\xi(t))+A(\xi(t))u(t)=v(t).
\end{equation}
Combining (\ref{sys-lin}) and (\ref{virtual}), we obtain the
linear (virtual) Input-Output mapping
\begin{equation}\label{linsys}
y^{(r)}(t)=v(t).
\end{equation}
Based on the linear system (\ref{linsys}), we propose the
stabilizing output feedback for the nominal system
(\ref{perturbed-model}) with $\Delta b(\xi)=0$, as
\begin{equation}\label{outputfeedback}
\begin{array}{l}
u_{nom}=A^{-1}(\xi)(v_{s}(t,\xi)-b(\xi)),\;v_{s}=(v_{s1},...,v_{sm})^{T}\\
v_{si}={y_{i}}^{(ri)}_{d}-K^{i}_{ri}({y_{i}}^{(ri-1)}-{y_{i}}^{(ri-1)}_{d})-...-K^{i}_{1}(y_{i}-{y_{i}}_{d}),\\
i\in\{1,...,m\}.
\end{array}
\end{equation}
Denoting the tracking error vector as
$e_{i}(t)=y_{i}(t)-{y_{i}}_{d}(t)$, we obtain the tracking error
dynamics
\begin{equation}\label{hurwitz_plys}
e_{i}^{(r_{i})}(t)+K^{i}_{r_{i}}e_{i}^{(r_{i}-1)}(t)+...+K^{i}_{1}e_{i}(t)=0,\;i=1,...,m,
\end{equation}
and by tuning the gains $K^{i}_{j},\;i=1,...,m,\;j=1,...,r_{i}$
such that all the polynomials in (\ref{hurwitz_plys}) are Hurwitz,
we obtain global asymptotic stability of the tracking errors
$e_{i}(t),\;i=1,...m$, to zero. To formalize this condition let us
state the following assumption.
\begin{assumption}\label{assumption_set_k} We assume that there exist a nonempty set
 $\mathcal{K}$ of gains $K^{i}_{j},\;i=1,...,m,\;j=1,...,r_{i}$,
 such that the polynomials (\ref{hurwitz_plys}) are Hurwitz.
 \end{assumption}
 \begin{remark}
Assumption \ref{assumption_set_k} is well know in the Input-Output
linearization control literature. It simply states that we can
find gains that stabilize the polynomials (\ref{hurwitz_plys}),
which can be done for example by pole placements.
 \end{remark}
Next, if we consider that $\Delta b(\xi)\neq 0$ in
(\ref{perturbed-model}), the global asymptotic stability of the
error dynamics will not be guarantied anymore due to the additive
error vector $\Delta b(\xi)$, we then choose to use Lyapunov
reconstruction technique (e.g. \cite{BL09-4}) to obtain a
controller ensuring practical stability of the tracking error.
This controller is presented in the following Theorem.
\begin{thm}\label{thm1}
Consider the system (\ref{systemeaffine1}) for any
$x_{0}\in\mathbb{R}^{n}$, under Assumptions 1, 2, 3, 4, 5 and 6,
with the feedback controller
\begin{equation}\label{outputfeedback_ftc}
\begin{array}{l}
u=A^{-1}(\xi)(v_{s}(t,\xi)-b(\xi))-A^{-1}(\xi)(\frac{\partial
V}{\partial z}_{ind})^{'}k\;d_{2}(e),\\k>0,\;\;
v_{s}=(v_{s1},...,v_{sm})^{T}\\
v_{si}={y_{i}}^{(ri)}_{d}-K^{i}_{ri}({y_{i}}^{(ri-1)}-{y_{i}}^{(ri-1)}_{d})-...-K^{i}_{1}(y_{i}-{y_{i}}_{d}).
\end{array}
\end{equation}
Where, $K^{i}_{j}\in\mathcal{K},\;j=1,...,ri,\;i=1,...,m$, and
$\frac{\partial V}{\partial z}_{ind}=(\frac{\partial V}{\partial
z(r1)},...,\frac{\partial V}{\partial z(rm)}),\;V=z^{T}Pz$, $P>0$
such that $P\tilde{A}+\tilde{A}^{T}P=-I$, with $\tilde{A}$ being
an $n\times n$ matrix defined as
\begin{equation}\label{tildea}
\tilde{A}=\left(\begin{array}{l}0,1,0,...................................,0\\0,0,1,0,................................,0\\\hspace{+2cm}\ddots\\-K^{1}_{1},...,-K^{1}_{r1},0,..................,0
\\\hspace{+2cm}\ddots\\0,...................,0,1,0,...........,0\\0,...................,0,0,1,...........,0\\\hspace{+2cm}\ddots\\0,................,0,-K^{m}_{1},...,-K^{m}_{rm}\end{array}\right),
\end{equation}
and
$z=(z^{1},...,z^{m})^{T},\;z^{i}=(e_{i},...,e_{i}^{r_{i}-1}),\;i=1,...,m$.
Then, the vector $z$ is uniformly bounded and reached the positive
invariant set $S=\{z\in\mathbb{R}^{n} | \;1-k\;|\frac{\partial
V}{\partial z}_{ind}|\geq 0\}$.
\end{thm}
{\it Proof:} The proof has been removed due to space constraints.
It will appear in a longer journal version of this work.
\subsection{Iterative tuning of the feedback gains}
In Theorem \ref{thm1}, we showed that the passive robust
controller (\ref{outputfeedback_ftc}) leads to bounded tracking
errors attracted to the invariant set $S$ for a given choice of
the feedback gains $K^{i}_{j},\;j=1,...,ri,\;i=1,...,m$. Next, to
iteratively tune the feedback gains of (\ref{outputfeedback_ftc}),
we define a desired cost function, and use a multi-variable
extremum seeking to iteratively auto-tune the gains and minimize
the defined cost function. We first denote the cost function to be
minimized as $Q(z(\beta))$
where $\beta$ represents the optimization variables vector,
defined as
\begin{equation}\label{beta}
\beta=[\delta K^{1}_{1},...,\delta K^{1}_{r1},...,\delta
K^{m}_{1},...,\delta K^{m}_{rm},\delta k]^{T}
\end{equation}
such that the updated feedback gains write as
\begin{equation}\label{estimated_para1}
\begin{array}{c}
K^{i}_{j}=K^{i}_{j-nominal}+\delta
K^{i}_{j},\;j=1,...ri,\;i=1,...,m.\\
k=k_{nominal}+\delta k,\;\; k_{nominal}>0
\end{array}
\end{equation}
where $K^{i}_{j-nominal},\;j=1,...ri,\;i=1,...,m$ are the nominal
initial values of the feedback gains chosen such that Assumption
(5) is satisfied.
\begin{remark}
The choice of the cost function $Q$ is not unique. For instance,
if the controller tracking performance at the time specific
instants $I t_{f},\;I=1,2,3...$ is important for the targeted
application (see the example presented in Section \ref{example}),
one can choose $Q$ as
\begin{equation}\label{cost_function1}
Q(z(\beta))=z^{T}(I t_{f})C_{1}z(I t_{f}),\;\;C_{1}>0
\end{equation}
If other performance needs to be optimized over a finite time
interval, for instance a combination of a tracking performance and
a control power performance, then one can choose for example the
cost function
\begin{equation}\label{cost_function2}
\begin{array}{l}
Q(z(\beta))=\int_{(I-1)t_{f}}^{I t_{f}}z^{T}(t)C_{1}z(t)dt+
\int_{(I-1)t_{f}}^{I t_{f}}u^{T}(t)C_{2}u(t)dt,\\I=1,2,3...,\;
C_{1},\; C_{2}>0
\end{array}
\end{equation}
The gains variation vector $\beta$ is then used to minimize the
cost function $Q$ over the iterations $I\in\{1,2,3,...\}$.
\end{remark}
Following multi-parametric extremum seeking theory \cite{AK02},
the variations of the gains are defined as
\begin{equation}\label{partial_para_estimat1}
\begin{array}{l}
\dot{x}_{K^{i}_{j}}=a_{K^{i}_{j}}sin(\omega_{K^{i}_{j}}t-\frac{\pi}{2})Q(z(\beta))\\
\delta{\hat K}^{i}_{j}(t)=x_{K^{i}_{j}}(t)+a_{K^{i}_{j}}sin(\omega_{K^{i}_{j}}t+\frac{\pi}{2}),\;j=1,...ri,\;i=1,...,m\\
\dot{x}_{k}=a_{k}sin(\omega_{k}t-\frac{\pi}{2})Q(z(\beta))\\
\delta\hat k(t)=x_{k}(t)+a_{k}sin(\omega_{k}t+\frac{\pi}{2}),\\
\end{array}
\end{equation}
where $a_{K^{i}_{j}},\;j=1,...ri,\;i=1,...,m,\;a_{k}$ are positive
tuning parameters, and
\begin{equation}\label{convergence_condition}
\begin{array}{l}
\omega_{1}+\omega_{2}\neq \omega_{3}, \;\text{for}
\;\omega_{1}\neq \omega_{2}\neq \omega_{3},\\\forall
\omega_{1},\omega_{2},\omega_{3}\in\{\omega
_{K^{i}_{j}},\omega_{k},\;j=1,...ri,\;i=1,...,m\},
\end{array}
\end{equation}
with $\omega_{i}>\omega^{*},\;\forall \omega_{i}\in\{\omega
_{K^{i}_{j}},\omega_{k},\;j=1,...ri,\;i=1,...,m\}$, $\omega^{*}$
large enough. \\To study the stability of the learning-based
controller, i.e. controller (\ref{outputfeedback_ftc}), with the
varying gains (\ref{estimated_para1}) and
(\ref{partial_para_estimat1}), we first need to introduce some
additional Assumptions.
\begin{assumption} \label{robustmesass11}
 We assume that the cost function $Q$ has a
local minimum at $\beta^*$.
\end{assumption}
\begin{assumption} \label{robustmesass21}
We consider that the initial gain vector $\beta$ is sufficiently
close to the optimal gain vector $\beta^{*}$.
\end{assumption}
\begin{assumption} \label{robustmesass31}
The cost function is analytic and its variation with respect to
the gains is bounded in the neighborhood of $\beta^{*}$, i.e.
$|\frac{\partial{Q}}{\partial
\beta}({\tilde{\beta}})|\leq\Theta_{2},\;\Theta_{2}>0,
\;\tilde{\beta}\in\mathcal{V}(\beta^{*})$, where
$\mathcal{V}(\beta^{*})$ denotes a compact neighborhood of
$\beta^{*}$.
\end{assumption}
 We can now state the following result.

\begin{thm}\label{thm2}
Consider the system (\ref{systemeaffine1}) for any
$x_{0}\in\mathbb{R}^{n}$, under Assumptions 1, 2, 3, 4, 5 and 6,
with the feedback controller
\begin{equation}\label{outputfeedback_ftc_learning}
\begin{array}{l}
u=A^{-1}(\xi)(v_{s}(t,\xi)-b(\xi))-A^{-1}(\xi)(\frac{\partial
V}{\partial z}_{ind})^{'}k(t)\;d_{2}(e),\\k>0, \;\;
v_{s}=(v_{s1},...,v_{sm})^{T},\\
v_{si}(t,\xi)=\hat{{y}_{i}}^{(ri)}_{d}-K^{i}_{ri}(t)({y_{i}}^{(ri-1)}-\hat{y_{i}}^{(ri-1)}_{d})-...\\-K^{i}_{1}(t)({y_{i}}-\hat{y_{i}}_{d}),\;
i=1,...,m
\end{array}
\end{equation}
Where, the state vector is reset following the resetting law
$x(It_{f})=x_{0},\;I\in\{1,2,...\}$,  the desired trajectory
vector is rest following
$\hat{y_{i}}_{d}(t)=y_{id}(t-(I-1)t_{f}),\;(I-1)t_{f}\leq
t<It_{f},\;I\in\{1,2,...\}$, and
$K^{i}_{j}(t)\in\mathcal{K},\;j=1,...,ri,\;i=1,...,m$  are
piecewise continues gains switched at each iteration $I$,
$I\in\{1,2,...\}$, following the update law
\begin{equation}\label{gains_update}
\begin{array}{l}
K^{i}_{j}(t)=K^{i}_{j-nominal}+\delta K^{i}_{j}(t)\\
\delta K^{i}_{j}(t)=\delta{\hat K}^{i}_{j}((I-1)t_{f}
),\;(I-1)t_{f}\leq t< It_{f},\\
k(t)=k_{nominal}+\delta k(t),\;\; k_{nominal}>0\\
\delta k(t)=\delta\hat k((I-1)t_{f}),\;(I-1)t_{f}\leq t<
It_{f},\;I=1,2,3...\\
\end{array}
\end{equation}
where $\delta{\hat K}^{i}_{j},\delta\hat k$ are given by
(\ref{partial_para_estimat1}), (\ref{convergence_condition}) and
whereas the rest of the coefficients are defined similarly to
Theorem \ref{thm1}. Then, the obtained closed-loop impulsive
time-dependent dynamic system (\ref{systemeaffine1}),
(\ref{partial_para_estimat1}), (\ref{convergence_condition}),
(\ref{outputfeedback_ftc_learning}) and (\ref{gains_update}), is
well posed, the tracking error $z$ is uniformly bounded, and is
steered at each iteration $I$ towards the positive invariant set
$S_{I}=\{z\in\mathbb{R}^{n} | \;1-k_{I}\;|\frac{\partial
V}{\partial z}_{ind}|\geq 0\}$, $k_{I}=\beta_{I}(n+1)$, where
$\beta_{I}$ is the value of $\beta$ at the $I$th iteration.
Furthermore, $ |Q(\beta(It_{f}))-Q(\beta^{*})|\leq
\Theta_{2}\big(\frac{\Theta_{1}}{\omega_{0}}+\sqrt{\sum\limits_{\small
i=1,...,m\;j=1,...,ri}{a_{K^{i}_{j}}}^{2}+{a_{k}}^{2}}\big),\;\Theta_{1},\;\Theta_{2}>0,\;\text{for
}\;I\rightarrow\infty $, where
$\omega_{0}=Max(\omega_{K^{1}_{1}},...,\omega_{K^{m}_{rm}},\omega_{k})$,
and $Q$ satisfies Assumptions \ref{robustmesass11},
\ref{robustmesass21} and \ref{robustmesass31}. Wherein, the vector
$\beta$ remains bounded over the iterations s.t.
$|\beta((I+1)t_{f})-\beta(It_{f})|\leq 0.5t_{f}
Max({a_{K^{1}_{1}}}^{2},...,{a_{K^{m}_{rm}}}^{2},a_{k}^{2})\Theta_{2}+t_{f}\omega_{0}\sqrt{\sum\limits_{\small
i=1,...,m\;j=1,...,ri}{a_{K^{i}_{j}}}^{2}+{a_{k}}^{2}},\;I\in\{1,2,...\}$,
and satisfies asymptotically the bound
$|\beta(It_{f})-\beta^{*}|\leq
\frac{\Theta_{1}}{\omega_{0}}+\sqrt{\sum\limits_{\small
i=1,...,m\;j=1,...,ri}{a_{K^{i}_{j}}}^{2}+{a_{k}}^{2}},\;\Theta_{1}>0,\;\text{for}\;I\rightarrow\infty
$.
\end{thm}
{\it Proof:} The proof has been removed due to space constraints.
It will appear in a longer journal version of this work.
\begin{remark}
In Theorem 2, we show that in each iteration $I$, the tracking
error vector $z$ is directed toward the invariant set $S_{I}$.
However, due to the finite time-interval length $t_{f}$ of each
iteration, we cannot guaranty that the vector $z$ enters $S_{I}$
in each iteration (unless we are in the trivial case where
$z_{0}\in S_{I}$). All what we guaranty is that the vector norm
$|z|$ starts from a bounded value $|z_{0}|$ and remains bounded
during the iterations with an upper-bound which can be estimated
as function of $|z_{0}|$ by using the bounds of the quadratic
Lyapunov functions $V_{I},\;I=1,2,...$, i.e. a uniform boundedness
result (\cite{HCN06}, p 6, def. 2.12).
\end{remark}
In the next section we propose to illustrate this approach on a
mechatronics system.
\section{The case of electromagnetic actuators}\label{example}
\begin{figure}
\begin{center}
\subfigure[Obtained armature position vs. reference trajectory -
Controller (\ref{example_u}) ]{
\includegraphics[width=.6\linewidth]{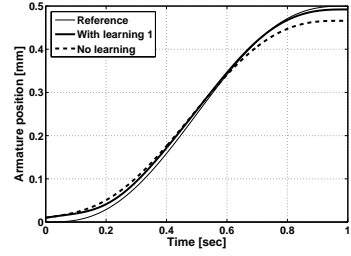}
\label{test1_x}} \subfigure[Obtained armature velocity vs.
reference trajectory - Controller (\ref{example_u})]{
\includegraphics[width=.6\linewidth]{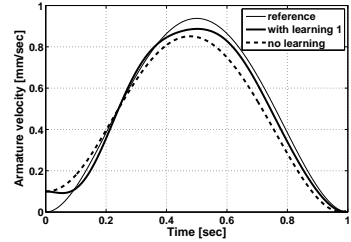}
\label{test1_v}}
\end{center}
\caption{Obtained outputs vs. reference trajectory - Controller
(\ref{example_u}) without learning (dashed line), with learning
(bold line)} \label{test1_x_v}
\end{figure}

\begin{figure}
\begin{center}
\subfigure[Obtained coil current]{
\includegraphics[width=.45\linewidth]{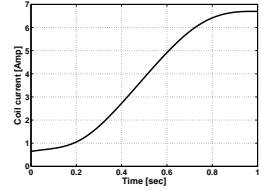}
\label{test1_i}} \subfigure[Control voltage]{
\includegraphics[width=.46\linewidth]{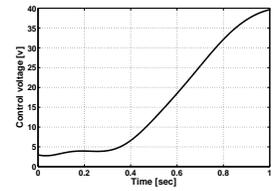}
\label{test1_u}}
\end{center}
\caption{Coil voltage and current - Controller (\ref{example_u})}
\label{test1_i_u}
\end{figure}

\begin{figure}
\begin{center}
\subfigure[Zoom at $t=t_{f}$ of the obtained armature velocity vs.
reference trajectory - Controller (\ref{example_u}) without
learning (dashed line), with learning (bold line)]{
\includegraphics[width=.6\linewidth]{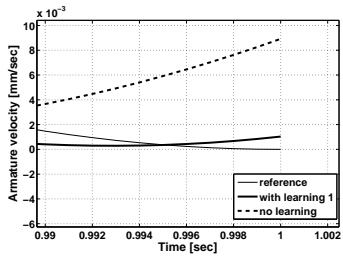}
\label{test1_v_zoom}} \subfigure[Cost function vs. learning
iterations]{
\includegraphics[width=.6\linewidth]{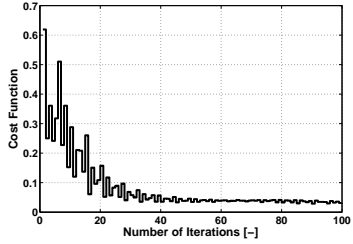}
\label{test1_Q}}
\end{center}
\caption{Impact velocity performance- Controller
(\ref{example_u})} \label{test1_Q_vzoom}
\end{figure}

\begin{figure}
\begin{center}
\subfigure[$K_{1}$ vs. learning iterations]{
\includegraphics[width=.45\linewidth]{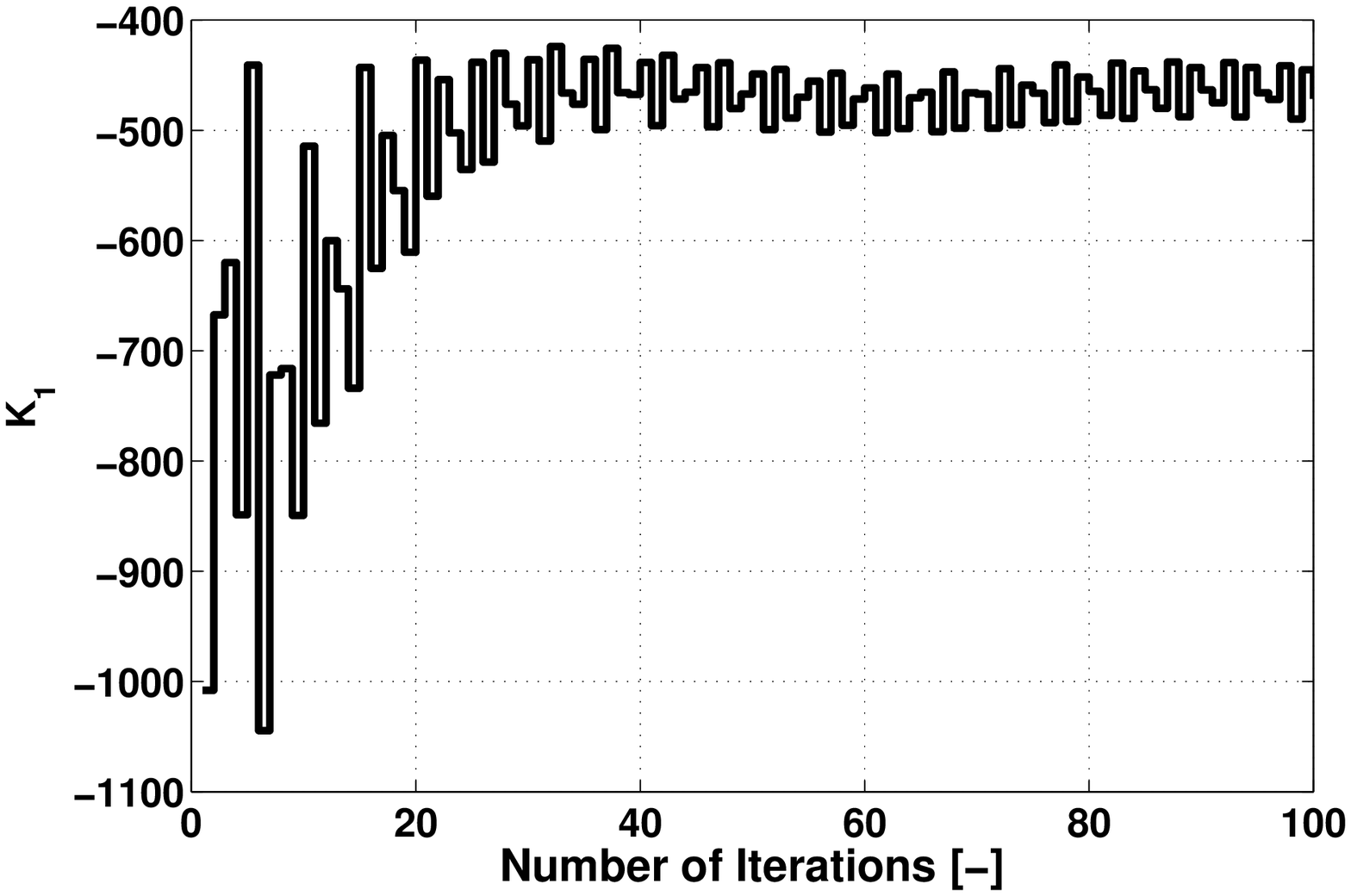}
\label{test1_k1}} \subfigure[$K_{2}$ vs. learning iterations]{
\includegraphics[width=.45\linewidth]{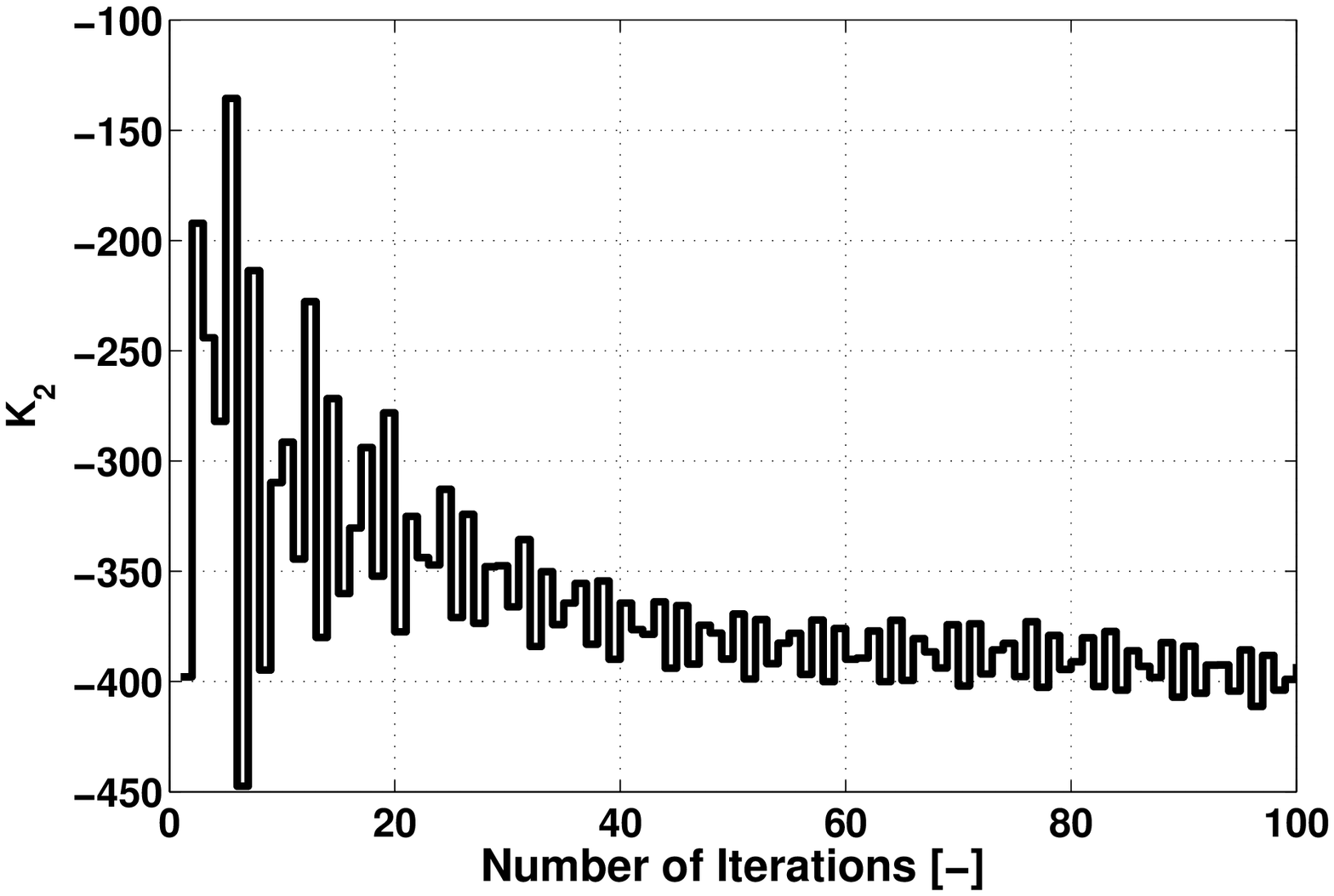}
\label{test1_k2}} \subfigure[$K_{3}$ vs. learning iterations]{
\includegraphics[width=.45\linewidth]{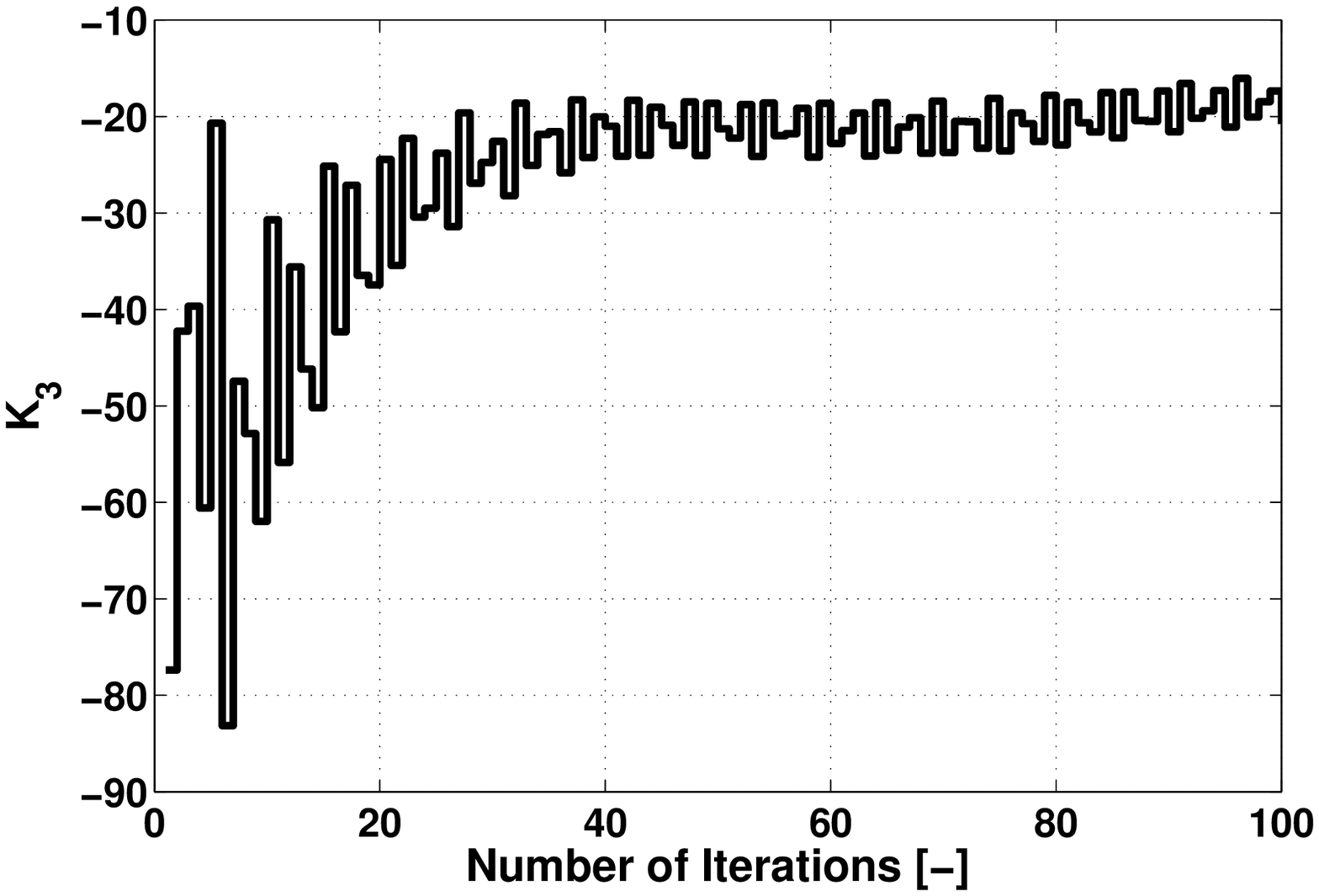}
\label{test1_k3}} \subfigure[$k$ vs. learning iterations]{
\includegraphics[width=.45\linewidth]{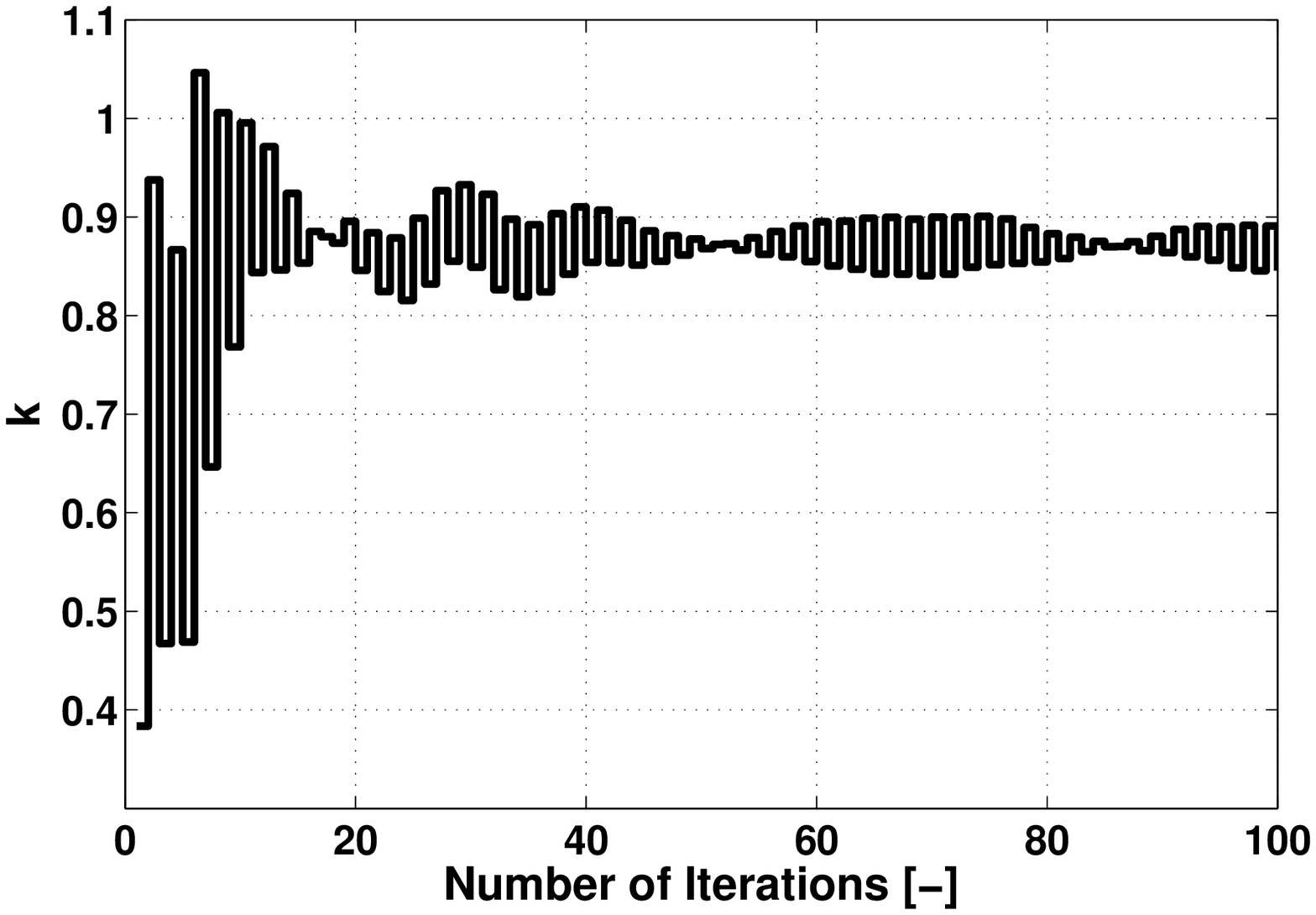}
\label{test1_k}} \vspace{-0cm}
\end{center}
\caption{Gains learning- Controller (\ref{example_u})}
\label{test1_Q_vzoom}
\end{figure}

We apply here the method presented above to the case of
electromagnetic actuators.\\
\underline{System modelling:}\label{model} Following
\cite{WSHKH00,PS04}, we consider the following nonlinear
 model for electromagnetic actuators
\begin{equation}
\begin{array}{l}
m\frac{d^2x_{a}}{dt^2}=k(x_0-x_{a})-\eta\frac{dx_{a}}{dt}-\frac{ai^2}{2(b+x_{a})^2}\\
u=Ri+\frac{a}{b+x_{a}}\frac{di}{dt}-\frac{ai}{(b+x_{a})^2}
\frac{dx_{a}}{dt},\;0\leq x_{a}\leq x_f,
\end{array}
\label{nonlin_model}
\end{equation}
where, $x_{a}$ represents the armature position physically
constrained between the initial position of the armature $0$, and
the maximal position of the armature $x_f$, $\frac{dx_{a}}{dt}$
represents the armature velocity, $m$ is the armature mass, $k$
the spring constant, $x_0$ the initial spring length, $\eta$ the
damping coefficient (assumed to be constant),
$\frac{ai^2}{2(b+x_{a})^2}$ represents the electromagnetic force
(EMF) generated by the coil, $a,b$ are two constant parameters of
the coil, $R$ the resistance of the coil, $L=\frac{a}{b+x_{a}}$
the coil inductance, $\frac{ai}{(b+x_{a})^2} \frac{dx_{a}}{dt}$
represents the back EMF. Finally, $i$ denotes the coil current,
$\frac{di}{dt}$ its time derivative and $u$ represents the control
voltage applied to the coil. In this model we do not consider the
saturation region of the flux linkage in the magnetic field
generated by the coil, since we assume a current and armature
motion ranges within the linear region of the flux.\\
{\underline{Passive robust controller:}} In this section we first
design a nonlinear passive robust control based on Theorem
\ref{thm1}.\\Follwoing Assumption \ref{traj_assumption}, we define
$x_{ref}$ a desired armature position trajectory, s.t.  $x_{ref}$
is a smooth (at least $C^2$) function satisfying the initial/final
constraints:
$x_{ref}(0)=0,\;x_{ref}(t_{f})=x_{f},\;\dot{x}_{ref}(0)=0,\;\dot{x}_{ref}(t_{f})=0$,
where $t_{f}$ is a desired finite motion time and $x_{f}$ is a
desired final position. We consider the dynamical system
(\ref{nonlin_model}) with bounded parametric uncertainties on the
spring coefficient $\delta k$, with $|\delta k|\leq \delta
k_{max}$, and the damping coefficient $\delta\eta$, with $|\delta
\eta|\leq \delta \eta_{max}$, such that $k=k_{nominal}+\delta k$,
$\eta=\eta_{nominal}+\delta\eta$, where
$k_{nominal},\;\eta_{nominal}$ are the nominal values of the
spring stiffness and the damping coefficient, respectively. If we
consider the state vector $x=(x_{a},\;\dot{x}_{a},\;i)^{'}$, and
the controlled output $x_{a}$, the uncertain model of
electromagnetic actuators can be written in the form of
(\ref{systemeaffine1}), as \vspace{-0cm} {\begin{equation}
\begin{array}{l}
\dot{x}=\left(
\begin{array}{l}
\dot{x}_{a}\\
\ddot{x}_{a}\\
\dot{i}
\end{array}\right)=\left(
\begin{array}{c}
x_{2}\\\frac{k_{nominal}}{m}(x_{0}-x_{1})-\frac{\eta_{nominal}}{m}x_{2}+...\\\hspace{+3cm}...-\frac{ax_{3}^{2}}{2(b+x_{1})^{2}}\\
-\frac{R(b+x_{1})}{a}x_{3}+\frac{x_{3}x_{2}}{b+x_{1}}
\end{array}
\right)\\+\left(
\begin{array}{c}
0\\\frac{\delta k}{m}(x_{0}-x_{1})+\frac{\delta\eta}{m}x_{2}\\
0
\end{array}\right)
+\left(
\begin{array}{c}
0\\0\\
\frac{b+x_{1}}{a}
\end{array}
\right)u\\
y=x_{1}.
\end{array}
\end{equation}}
Assumption 1 is clearly satisfied over a nonempty bounded set $X$,
as for Assumption 2, it is straightforward to check that if we
compute the third time-derivative of the output $x_{a}$, the
control variable $u$ appears in a nonsingular expression, which
implies that $r=n=3$. Assumption 3 is also satisfied since
$|\Delta f(x)|\leq \frac{\delta
k_{max}}{m}|x_{0}-x_{1}|+\frac{\delta\eta_{max}}{m}|x_{2}|$.\\Next,
following the Input-Output linearization method, we can write
\vspace{-0cm}
\begin{equation}\label{linearization_equation}
\begin{array}{l}
y^{(3)}=x_{a}^{(3)}=-\frac{k_{nominal}}{m}\dot{x}_{a}-\frac{\eta_{nominal}}{m}\ddot{x}_{a}+\frac{Ri^{2}}{(b+x_{a})m}-\\\frac{\delta
k}{m}\dot{x}_{a}-\frac{\delta\eta}{m}\ddot{x}_{a}-\frac{i}{m(b+x_{a})}u,
\end{array}
\end{equation}
which is of the form of equation (\ref{perturbed-model}), with
$A=-\frac{i}{m(b+x_{a})},\;b=-\frac{k_{nominal}}{m}\dot{x}_{a}-\frac{\eta_{nominal}}{m}\ddot{x}_{a}+\frac{Ri^{2}}{(b+x_{a})m}$,
and the additive uncertainty term $\Delta b=-\frac{\delta
k}{m}\dot{x}_{a}-\frac{\delta\eta}{m}\ddot{x}_{a} $, such that
$|\Delta b|\leq \frac{\delta
k_{max}}{m}|\dot{x}_{a}|+\frac{\delta\eta_{max}}{m}|\ddot{x}_{a}|
=d_{2}(x_{a},\dot{x}_{a})$. Let us define the tracking error
vector $\mathbf{z}:=(z_1,\; z_2,\;
z_3)'=(x_{a}-x_{ref},\;\dot{x}_{a}-\dot{x}_{ref},\;
\ddot{x}_{a}-\ddot{x}_{ref})'$, where
$\dot{x}_{ref}=\frac{dx_{ref}(t)}{dt}$, and
$\ddot{x}_{ref}=\frac{d^{2} x_{ref}(t)}{dt^{2}}$. Next, using
Theorem \ref{thm1}, we can write the following robust passive
controller
\begin{equation}\label{example_u}
\begin{array}{l}
u=-\frac{m(b+x_{a})}{i}(v_{s}+\frac{k_{nominal}}{m}\dot{x}_{a}+\frac{\eta_{nominal}}{m}\ddot{x}_{a}-\frac{Ri^{2}}{(b+x_{a})m})
+\\\frac{m(b+x_{a})}{i}\frac{\partial V}{\partial
z_{3}}k(\frac{\delta
k_{max}}{m}|\dot{x}_{a}|+\frac{\delta\eta_{max}}{m}|\ddot{x}_{a}|),\;k>0\\
v_{s}=x_{ref}^{(3)}(t)+K_{3}(x_{a}^{(2)}-x_{ref}^{(2)}(t))+K_{2}(x_{a}^{(1)}-x_{ref}^{(1)}(t))\\+K_{1}(x_{a}-x_{ref}(t)),\;K_{i}<0,
i=1,2,3.
\end{array}
\end{equation}
Where, $V=z^{T}Pz$, $P>0$ solution of the equation
$P\tilde{\tilde{A}}+\tilde{\tilde{A}}^{T}P=-I$, with
\begin{equation}\label{tildea_example}
\tilde{\tilde{A}}=\left(\begin{array}{ccc}0&1&0\\0&0&1\\K_{1}&K_{2}&K_{3}\end{array}\right),
\end{equation}
where $K_{1},\;K_{2},\;K_{3}$ are chosen such that
$\tilde{\tilde{A}}$ is Hurwitz.\\
\underline{Learning-based auto-tuning of the controller gains:} We
use now the results of Theorem \ref{thm2}, to iteratively
auto-tune the feedback gains of the controller (\ref{example_u}).
 Considering a cyclic behavior of the actuator with each iteration happening over a time interval of length $t_{f}$, following (\ref{cost_function1}) we define the cost function as
\begin{equation}\label{cost_function_example}
Q(z(\beta))=C_{1}z_1(It_{f})^2+C_{2}z_2(It_{f})^2+C_{3}z_{3}(It_{f})^2,
\end{equation}
where $I=1,2,3...$ is the number of iterations,
$C_{1},\;C_{2}>0,\;C_{3}>0$, and $\beta=(\delta K_{1},\;\delta
K_{2},\;\delta K_{3},\;\delta k)'$, such as the feedback gains
write as
\begin{equation}\label{estimated_para}
\begin{array}{l}
K_{1}=K_{1_{nominal}}+\delta K_{1}\\
K_{2}=K_{2_{nominal}}+\delta K_{2}\\
K_{3}=K_{3_{nominal}}+\delta K_{3}\\
k={k}_{nominal}+\delta {k},
\end{array}
\end{equation}
where
$K_{1_{nominal}},\;K_{2_{nominal}},\;K_{3_{nominal}},\;{k}_{nominal}$
are the nominal initial values of the feedback gains in
(\ref{example_u}).\\Folowing (\ref{partial_para_estimat1}),
(\ref{convergence_condition}), and (\ref{gains_update}) the
variations of the estimated gains are given by
\begin{equation}\label{partial_para_estimat}
\begin{array}{l}
\dot{x}_{K_{1}}=a_{K_{1}}sin(\omega_{1}t-\frac{\pi}{2})Q(z(\beta))\\
\delta\hat{K}_{1}(t)=x_{K_{1}}(t)+a_{{K}_{1}}sin(\omega_{1}t+\frac{\pi}{2})\\
\dot{x}_{K_{2}}=a_{K_{2}}sin(\omega_{2}t-\frac{\pi}{2})Q(z(\beta))\\
\delta\hat{K}_{2}(t)=x_{{K}_{2}}(t)+a_{{K}_{2}}sin(\omega_{2}t+\frac{\pi}{2})\\
\dot{x}_{{K}_{3}}=a_{K_{3}}sin(\omega_{3}t-\frac{\pi}{2})Q(z(\beta))\\
\delta\hat{K}_{3}(t)=x_{{K}_{3}}(t)+a_{{K}_{3}}sin(\omega_{3}t+\frac{\pi}{2})\\
\dot{x}_{k}=a_{k}sin(\omega_{4}t-\frac{\pi}{2})Q(z(\beta))\\
\delta\hat{k}(t)=x_{k}(t)+a_{k}sin(\omega_{4}t+\frac{\pi}{2})\\
\delta K_{j}(t)=\delta\hat K_{j}((I-1)t_{f}),\;(I-1)t_{f}\leq t< I
t_{f},\\j\in\{1,2,3\},\;I=1,2,3...\\
\delta k(t)=\delta\hat k((I-1)t_{f}),\;(I-1)t_{f}\leq t< It_{f},
\;I=1,2,3...
\end{array}
\end{equation}
where $a_{K_{1}},\;a_{{K}_{2}},\;a_{{K}_{3}},\;a_{k}$ are positive
and $
\omega_{p}+\omega_{q}\neq \omega_{r},\;p,q,r\in\{1,2,3,4\},\;\text{for}\;p\neq q\neq r.\\
$\\
\underline{Simulation results:}\label{simulation} We show here the
behavior of the proposed approach on the electromagnetic actuator
example presented in \cite{KK010}, where the model
(\ref{nonlin_model}) is used with the numerical values of Table
\ref{table_1}.
\begin{table}
\begin{center}
\begin{tabular}{|c|c|}
\hline {\bf Parameter}  & {\bf Value}\\\hline
$m$ & $0.27\; [kg]$\\
$R$  & $6\; [\Omega]$\\
$\eta$  & $7.53\; [kg/sec]$\\
$x_{0}$  & $8\; [mm]$\\
$k$  & $158\; [N/mm]$\\
$a$  & $14.96\times 10^{-6}\;[Nm^{2}/A^{2}]$\\
$b$  & $ 4\times 10^{-5}\; [m]$\\
\hline
\end{tabular}\caption{Numerical values of the mechanical parameters}\vspace{-1cm}
\label{table_1}
\end{center}
\end{table}
 The desired trajectory has been selected as the
$5th$ order polynomial
$x_{ref}(t)=\sum_{i=0}^{5}a_{i}(t/t_{f})^{i}$, where the $a_{i}$'s
have been computed to satisfy the boundary constraints
$x_{ref}(0)=0,x_{ref}(t_{f})=x_{f},\dot{x}_{ref}(0)=\dot{x}_{ref}(t_{f})=0,\ddot{x}_{ref}(0)=\ddot{x}_{ref}(t_{f})=0$,
with $t_{f}=1\;sec$, $x_{f}=0.5\;mm$.\\
Furthermore, to make the simulation case more challenging we
assume an initial error both on the position and the velocity
$z_{1}(0)=0.01\;mm$, $z_{2}(0)=0.1\;mm/sec$. Note that these
values may seem small, but for this type of actuators it is
usually the case that the armature starts form a predefined static
position constrained mechanically, so we know that the initial
velocity is zero and we know in advance very precisely the initial
position of the armature. However, we want to show the
performances of the controller on some challenging cases. We also
select the nominal feedback gains
$K_{1}=-500,\;K_{2}=-125,\;K_{3}=-26,\;k=1$, satisfying Assumption
5. In this test we compare the performances of the passive robust
controller (\ref{example_u}) with the fixed nominal gains, to the
learning controller (\ref{example_u}),(\ref{estimated_para}),
(\ref{partial_para_estimat}), which was implemented with the cost
function (\ref{cost_function_example}), where
$C_{1}=500,\;C_{2}=500,\;C_{3}=10$, and the learning coefficients
for each feedback gain are $\omega_{1}=7.5\;rad/sec$,
$\omega_{2}=5.3\;rad/sec$, $\omega_{3}=5.1\;rad/sec$,
$\omega_{4}=6.1\;rad/sec$. We point out here that to accelerate
the learning convergence rate, which is related to the choice of
the coefficients $a_{K_{i}},\;i=1,2,3,\;a_{k}$ , e.g.
\cite{TNM08}, we have chosen to use a varying amplitude for the
coefficients. Indeed, it is well know , e.g. \cite{MMB09}, that
choosing varying coefficients, which start with a high value to
accelerate the search initially and then are tuned down when the
cost function becomes smaller, accelerates the learning and
achieves a convergence to a tighter neighborhood of the local
optimum (due to decrease of the dither amplitudes). To implement
this idea, we simply use piece-wise constant coefficients as
follows: $a_{{K}_{1}}=200$, $a_{{K}_{2}}=120$, $a_{{K}_{3}}=20$,
$a_{{k}}=0.2$, initially and then tuned them down to
$a_{{K}_{1}}=200Q(1)/2$, $a_{{K}_{2}}=120Q(1)/2$,
$a_{{K}_{3}}=20Q(1)/2$, $a_{{k}}=0.2Q(1)/2$, when $Q\leq Q(1)/2$
and then to $a_{{K}_{1}}=200Q(1)/3$, $a_{{K}_{2}}=120Q(1)/3$,
$a_{{K}_{3}}=2Q(1)/3$, $a_{{k}}=0.2Q(1)/3$, when $Q\leq Q(1)/3$,
where $Q(1)$ denotes the value of the cost function at the first
iteration. We show on figures \ref{test1_x}, \ref{test1_v} the
performance of the position and the velocity tracking, with and
without the learning algorithm. We see clearly the effect of the
learning algorithm that makes the landing velocity closer to the
desired zero landing velocity as shown on figure
\ref{test1_v_zoom}. The associated coil current and voltage
signals are also reposted on figures \ref{test1_i} and
\ref{test1_u}, respectively. It is worth mentioning here that the
optimized performance in this example is focused mainly on the
impact point, i.e. the position and velocity of the armature at
$t=t_{f}$, this is why we choose a cost function as
(\ref{cost_function_example}) instead of a cost function based on
the integral of the tracking error. We also report on figure
\ref{test1_Q}, the cost function value along the learning
iterations. We see a clear decrease of the cost function which
reaches a local optimum after about $40$ iterations. We point out
here that the transient behavior of the cost function which
oscillates with relative large amplitude is due to the choice of
learning amplitudes $a_{K_{i}}$'s , which we choose to initiate at
high values to accelerate the learning process. We can obtain much
lower excursion amplitudes during the transient behavior at the
expense of the convergence speed, by choosing smaller learning
amplitudes. We also report the learned feedback gains on figures
\ref{test1_k1}, \ref{test1_k2}, \ref{test1_k3}, and \ref{test1_k},
respectively. They also show a trend of convergence, with final
oscillations around the convergence point. The excursion of these
oscillations can be easily tuned by the tuning of the learning
coefficients $a_{K_{i}},\;i=1,2,3,4$.


\section{Conclusion}\label{conclusion}
In this work we have studied the problem of iterative feedback
gains tuning for Input-Output linearization with static state
feedback. We first used Input-Output linearization with static
state feedback method and `robustified' it with respect to bounded
additive model uncertainties, using Lyapunov reconstruction
techniques, to ensure uniform boundedness of a tracking error
vector. Secondly, we complemented the Input-Output linearization
controller with a model-free learning algorithm to iteratively
auto-tune the control feedback gains and optimize a desired
performance of the system. The learning algorithm used here is
based on multi-parametric extremum seeking theory. The full
controller, i.e. the learning algorithm together with the passive
robust controller forms an iterative gains auto-tuning
Input-Output linearization controller. We have reported some
numerical results obtained on an electromagnetic actuators
example. Future investigations will focuss on improving the
convergence rate by using different MES algorithms with
semi-global convergence properties, e.g. \cite{TNM06,MTNM11,S13},
extending this work to different type of model-free learning
algorithms, e.g. reinforcement learning algorithms, and comparing
the learning algorithms in terms of their convergence rate and
achievable optimal performances.


\begin{thebibliography}{10}
\providecommand{\url}[1]{#1}
\def\UrlFont{\rmfamily}
\providecommand{\newblock}{\relax}
\providecommand{\bibinfo}[2]{#2}
\providecommand\BIBentrySTDinterwordspacing{\spaceskip=0pt\relax}
\providecommand\BIBentryALTinterwordstretchfactor{4}
\providecommand\BIBentryALTinterwordspacing{\spaceskip=\fontdimen2\font
plus \BIBentryALTinterwordstretchfactor\fontdimen3\font minus
  \fontdimen4\font\relax}
\providecommand\BIBforeignlanguage[2]{{%
\expandafter\ifx\csname l@#1\endcsname\relax
\typeout{** WARNING: IEEEtran.bst: No hyphenation pattern has been}%
\typeout{** loaded for the language `#1'. Using the pattern for}%
\typeout{** the default language instead.}%
\else \language=\csname l@#1\endcsname \fi #2}}

\bibitem{I89}
A.~Isidori, \emph{Nonlinear Control Systems}, 2nd~ed., ser.
Communications and
  Control Engineering Series.\hskip 1em plus 0.5em minus 0.4em\relax
  {S}pringer-{V}erlag, 1989.

\bibitem{FK08}
L.~Freidovich and H.~Khalil, ``Performance recovery of
feedback-linearization
  based designs,'' \emph{IEEE, Transactions on Automatic Control}, vol.~53,
  no.~10, pp. 2324--2334, November 2008.

\bibitem{CW95}
Y.-S. Chou and W.~Wu, ``Robust nonlinear control associating
robust feedback
  linearization and {H}$_{\infty}$ control,'' \emph{Chemical Enginnering
  Science}, vol.~50, no.~9, pp. 1429--1439, 1995.

\bibitem{PD11}
C.~Pop and E.~Dulf, \emph{Recent advances in robust control- Novel
approaches
  and design methods}, intech~ed., 2011, ch. Robust feedback linearization
  control for reference tracking and disturbance rejection in nonlinear
  systems, pp. 274--290.

\bibitem{FBPG06}
A.~L.~D. Franco, H.~Bourlès, E.~R. de~Pieri, and H.~Guillard,
``Robust
  nonlinear control associating robust feedback linearization and
  {H}$_{\infty}$ control,'' \emph{IEEE, Transactions on Automatic Control},
  vol.~51, no.~7, pp. 1200--1207, November 2006.

\bibitem{LGMBT03}
O.~Lequin, M.~Gevers, M.~Mossberg, E.~Bosmans, and L.~Triest,
``Iterative
  feedback tuning of {PID} parameters: comparison with classical tuning
  rules,'' \emph{Control Engineering Practice}, vol.~11, no.~9, pp. 1023 --
  1033, 2003.

\bibitem{H02}
\BIBentryALTinterwordspacing H.~Hjalmarsson, ``Iterative feedback
tuning—an overview,'' \emph{International
  Journal of Adaptive Control and Signal Processing}, vol.~16, no.~5, pp.
  373--395, 2002. [Online]. Available: \url{http://dx.doi.org/10.1002/acs.714}
\BIBentrySTDinterwordspacing

\bibitem{KK06}
N.~Killingsworth and M.~Kristic, ``{PID} tunning using extremum
seeking,''
  \emph{IEEE Control Systems Magazine}, pp. 1429--1439, 2006.

\bibitem{KRK06}
L.~Koszalka, R.~Rudek, and I.~Pozniak-Koszalka, ``An idea of using
  reinforcement learning in adaptive control systems,'' in \emph{Networking,
  International Conference on Systems and International Conference on Mobile
  Communications and Learning Technologies, 2006. ICN/ICONS/MCL 2006.
  International Conference on}, April 2006, pp. 190--196.

\bibitem{benosmanatincacc13}
M.~Benosman and G.~Atinc, ``Multi-parametric extremum
seeking-based learning
  control for electromagnetic actuators,'' in \emph{American Control
  Conference}, 2013, pp. 1917--1922.

\bibitem{K96}
H.~Khalil, \emph{Nonlinear systems}, 2nd~ed.\hskip 1em plus 0.5em
minus
  0.4em\relax New York Macmillan, 1996.

\bibitem{BL09-4}
M.~Benosman and K.-Y. Lum, ``Passive actuators' fault tolerant
control for
  affine nonlinear systems,'' \emph{IEEE, Transactions on Control Systems
  Technology}, vol.~18, no.~1, pp. 152--163, January 2010.

\bibitem{HCN06}
W.~M. Haddad, V.~Chellaboind, and S.~G. Nersesov, \emph{Impulsive
and Hybrid
  Dynamical Systems: Stability, Dissipativity, and Control}.\hskip 1em plus
  0.5em minus 0.4em\relax Princeton University Press, Princeton, 2006.

\bibitem{EO92}
H.~Elmali and N.~Olgac, ``Robust output tracking control of
nonlinear mimo
  systems via sliding mode technique,'' \emph{Automatica}, vol.~28, no.~1, pp.
  145--151, 1992.

\bibitem{AK02}
K.~B. Ariyur and M.~Krstic, ``Multivariable extremum seeking
feedback: Analysis
  and design,'' in \emph{Proc. of the Mathematical Theory of Networks and
  Systems}, South Bend, IN, August 2002.

\bibitem{WSHKH00}
Y.~Wang, A.~Stefanopoulou, M.~Haghgooie, I.~Kolmanovsky, and
M.~Hammoud,
  ``Modelling of an electromechanical valve actuator for a camless engine,'' in
  \emph{5th International Symposium on Advanced Vehicle Control}, 2000, number
  93.

\bibitem{PS04}
K.~Peterson and A.~Stefanopoulou, ``Extremum seeking control for
soft landing
  of electromechanical valve actuator,'' \emph{Automatica}, vol.~40, pp.
  1063--1069, 2004.

\bibitem{KK010}
N.~Kahveci and I.~Kolmanovsky, ``Control design for
electromagnetic actuators
  based on backstepping and landing reference governor,'' in \emph{5th IFAC
  Symposium on Mechatronic Systems}, Cambridge, September 2010, pp. 393--398.

\bibitem{TNM08}
Y.~Tan, D.~Nesic, and I.~Mareels, ``On the dither choice in
extremum seeking
  control,'' \emph{Automatica}, no.~44, pp. 1446--1450, 2008.

\bibitem{MMB09}
W.~Moase, C.~Manzie, and M.~Brear, ``Newton-like extremum seeking
part {I}:
  Theory,'' in \emph{IEEE, Conference on Decision and Control}, December 2009,
  pp. 3839--3844.

\bibitem{TNM06}
Y.~Tan, D.~Nesic, and I.~Mareels, ``On non-local stability
properties of
  extremum seeking control,'' \emph{Automatica}, no.~42, pp. 889--903, 2006.

\bibitem{MTNM11}
W.~Noase, Y.~Tan, D.~Nesic, and C.~Manzie, ``Non-local stability
of a
  multi-variable extremum-seeking scheme,'' in \emph{IEEE, Australian Control
  Conference}, November 2011, pp. 38--43.

\bibitem{S13}
A.~Scheinker, ``Simultaneous stabilization of and optimization of
unkown
  time-varying systems,'' in \emph{American Control Conference}, June 2013, pp.
  2643--2648.

\end{thebibliography}
\end{document}